\documentclass[twocolumn,showpacs,preprintnumbers,,superscriptaddress]{revtex4}
\usepackage{graphicx}
\usepackage{epsf}
\usepackage{dcolumn}
\usepackage{bm}
\usepackage{amsmath}

\begin{document}

\title{Spin excitations in the antiferromagnet NaNiO$_{2}$}

\author{S. de Brion}
\affiliation{Grenoble High Magnetic Field Laboratory, CNRS, BP
166, 38042 Grenoble, France}
\author{C. Darie}
\affiliation{Laboratoire de Cristallographie, CNRS, BP 166, 38042
Grenoble, France}
\author{M. Holzapfel}
\affiliation{Grenoble High Magnetic Field Laboratory, CNRS, BP
166, 38042 Grenoble, France} \affiliation{present address:
Paul-Scherrer-Institut CH-5232 Villigen PSI, Switzerland}
\author{D. Talbayev}
\affiliation{Department of Physics, Stony Brook University, NY
11794-3800, USA}
\author{L. Mih\'aly}
\affiliation{Department of Physics, Stony Brook University, NY
11794-3800, USA} \affiliation{Electron Transport Research Group of
the Hungarian Academy of Science and Department of Physics,
Budapest University of Technology and Economics, 1111 Budapest,
Hungary}
\author{F. Simon}
\affiliation{Budapest University of Technology and Economics,
Institute of Physics and Solids in Magnetic Fields Research Group
of the Hungarian Academy of Sciences, P. O. Box 91, H-1521
Budapest, Hungary}
\author{A. J\'anossy}
\affiliation{Budapest University of Technology and Economics,
Institute of Physics and Solids in Magnetic Fields Research Group
of the Hungarian Academy of Sciences, P. O. Box 91, H-1521
Budapest, Hungary}
\author{G. Chouteau}
\affiliation{Grenoble High Magnetic Field Laboratory, CNRS, BP
166, 38042 Grenoble, France}

\date{\today}

\begin{abstract}
In NaNiO$_{2}$ , Ni$^{3+}$ ions form a quasi two dimensional
triangular lattice of $S=1/2$ spins. The magnetic order observed
below 20K has been described as an A type antiferromagnet with
ferromagnetic layers weakly coupled antiferromagnetically. We
studied the magnetic excitations with the electron spin resonance
for frequencies 1-20~cm$^{-1}$,  in magnetic fields up to 14 T. The
bulk of the results are interpreted in terms of a phenomenological
model involving bi-axial anisotropy for the spins: a strong
easy-plane term, and a weaker anisotropy within the plane. The
direction of the easy plane is constrained by the collective
Jahn-Teller distortion occurring in this material at 480~K.
\end{abstract}

 \maketitle

\section{\label{sec:level1}Introduction}

A two-dimensional triangular network of magnetic ions interacting
via an antiferromagnetic interaction is a well known geometrically
frustrated system where unconventional magnetic properties are
expected \cite{frustration}. Usually, a long range magnetic order
occurs at low enough temperature. For instance, in XCl$_{2}$ with
X=Cr, Br, or in ACrO$_{2}$ with A=Li, Ag, Cu, or CsCuCl$_{3}$ the
magnetic order is based on a  120$^{\circ}$ spin structure on the
triangles. All these compounds have an  easy plane or easy axis
anisotropy together with  Heisenberg type antiferromagnetic
interactions. In other compounds, no magnetic order was detected
so far (NaCrO$_{2}$, KCrO$_{2}$, NaTiO$_{2}$, LiNiO$_{2}$). The
possibility of the orbital order competing with the spin order
makes LiNiO$_{2}$ particularly interesting.  In this compounds the
Ni$^{3+}$ ions have a spin $S=1/2$ and the orbital occupation is
 doubly degenerate: the $e_g$ orbitals, $\mid 3z^{2}-r^{2}>$
and $\mid x^{2}-y^{2}>$, have the same energy unless the oxygen
octahedron around the magnetic ion becomes elongated or compressed
due to the Jahn-Teller effect.

Surprisingly, no orbital order has been observed in LiNiO$_{2}$.
The absence of both orbital and magnetic order in this compound
has been the subject of intense debate lately, both experimentally
\cite{LiNiO2 exp} and theoretically \cite{LiNiO2 theo}. In
particular, there is still a controversy on what are the relevant
magnetic interactions within the triangular planes and to what
extend this compound is magnetically frustrated. The comparison
with the isomorphic NaNiO$_{2}$ is aimed at elucidating this
unconventional behavior. In NaNiO$_{2}$ a ferro-distortive orbital
order (a collective Jahn-Teller distortion) is observed below
480~K \cite{NaNiO2 ferro orbital} and a long range
antiferromagnetic order appears below 20~K \cite{NaNiO2 AF}. This
magnetic order was first described as an A type antiferromagnet
with ferromagnetic planes coupled antiferromagnetically
\cite{NaNiO2 AF}\cite{NaNiO2 Bongers}. The magnetic superlattice
has indeed been observed recently in neutron diffraction
measurements \cite{NaNiO2 neutron}. However, the description of
the magnetic system with just two magnetic interactions (an
antiferromagnetic $J_{AF}$ between subsequent NiO planes and a
ferromagnetic $J_{F}$ within the planes) fails to describe the
whole magnetic behavior, in particular the presence of three
characteristic fields observed in the magnetization curve
\cite{LiNaNiO}. The direction of the spins measured by neutron
diffraction is also unusual: they point toward the center of one
of the triangles in the oxygen octahedron surrounding each Ni ion,
at $100^{\circ}$ from the Ni plane \cite{NaNiO2 neutron}.
Moreover, a recent inelastic neutron study \cite{lewis2005} has
shown that a gap of $\simeq$0.7~meV is present in the spin
excitation spectrum.

We have performed Electron Spin Resonance measurements in magnetic
fields up to 14~T on powder samples of NaNiO$_{2}$.  We adopted
and extended the model used to describe the magnon spectrum
\cite{lewis2005}.  We find that the low field behavior of the
system is characteristic of an "easy-plane" magnet, with a small
anisotropy within the plane.  The spin-flop transition is assigned
to this latter anisotropy.  At high fields all spins are aligned
parallel and the saturation effects dominate the behavior.
Although the model described here is generally successful, we also
observed a spin resonance mode that remains unexplained. The
possible implications for the magnetic interactions in LiNiO$_{2}$
are discussed.

\section{Experiments}

The NaNiO$_{2}$ powdered sample was obtained following the
procedure described elsewhere \cite{NaNiO2 ferro orbital}.  The
monoclinic C2/m crystal structure was checked by X-ray powder
diffraction. The system is quasi two-dimensional: The Ni-Ni
interplane distance is large, 5.568~$\AA$. Below the Jahn-Teller
transition at 480~K the Ni$^{3+}$ ions are arranged in a slightly
distorted triangular network, with one short length and two longer
ones (at room temperature, the Ni-Ni distances are 2.84~$\AA$ and
3.01~$\AA$ respectively compared to 2.96~$\AA$ at 565~K)
\cite{NaNiO2 ferro orbital}. Also, the oxygen octahedra
surrounding the nickel ions are elongated. Such a distortion
favors the $\mid 3z^{2}-r^{2}>$ orbitals.  Indeed, the ESR spectra
at 200K have the anisotropic shape typical of the $\mid
3z^{2}-r^{2}>$ orbitals, with $g^\parallel=2.03$ and
$g^\perp=2.28$ for fields parallel and perpendicular to $z$,
respectively \cite{NaNiO2 ferro orbital}.

The magnetic behavior of our NaNiO$_2$ sample was the same as the
one reported previously \cite{LiNaNiO}. The magnetic susceptibility
exhibits a Curie-Weiss behavior above 100~K. The Curie-Weiss
temperature of $T_{\rm CW}$=36~K reflects the predominance of
ferromagnetic interactions. This is a noticeable difference compared
to frustrated triangular compounds such as LiCrO$_2$. The effective
magnetic moment (1.85 $\mu$$ _{B}$) is in agreement with the $\mid
3z^{2}-r^{2}>$ orbital configuration
 for the
$e_g$  electron in the low spin state of Ni$^{3+}$ (spin S=1/2,
average gyromagnetic factor $g_{av}=2.14$). The N\'{e}el
temperature, observed as a peak in the susceptibility, is $T_{\rm
N}$=20~K. The magnetization curve in the ordered phase at 4~K
presents two kinks at $H_{\rm C0}$=1.8~T, $H_{\rm C1}$= 8~T,
previously assigned to spin flop transitions, and the
magnetization is fully saturated at $H_{\rm {sat}}$= 13~T
\cite{LiNaNiO}.

\begin{figure}

  \includegraphics[width=7cm]{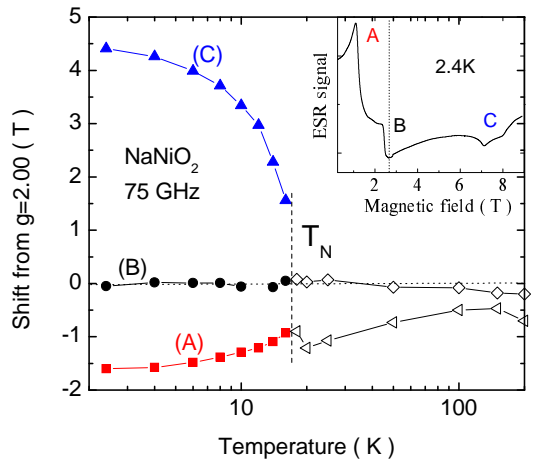}
  \caption{Shift of the resonance lines position from the paramagnetic line at 75 GHz as a function of temperature. Insert: typical ESR spectrum as a function of magnetic field
recorded at 2.4 K.The dotted line corresponds to the paramagnetic
line at g=2 .} \label{Figure 1}
\end{figure}

The Electron Spin Resonance measurements (ESR) were performed
using two different techniques. Classical high frequency, high
magnetic field ESR was used at fixed frequencies using Gunn diode
oscillators and multipliers at 35, 75, 150~ and 225~GHz, combined
with a 9~T magnet at the  Budapest University of Technology and
Economics. Previous measurements at the Grenoble High Magnetic
Field laboratory have been described in Ref. \onlinecite{NaNiO2
AF}.

Other measurements were performed by a novel method \cite{diyar,
Mihaly LaMnO3}.  We used the far infrared facilities of the
National Synchrotron Light Source in Brookhaven National
Laboratory, at the U12 IR beamline, including a 14~T
superconducting magnet (Oxford Instruments), and a Fourier
Transform Infrared Spectrometer (Sciencetech, SPS 200). We will
present results obtained in the frequency range 3~cm$^{-1}$ -
20~cm$^{-1}$ (90~GHz - 600~GHz) at 4 K. The powdered sample was in
a disk-shaped teflon container of 5~mm diameter and 1~mm
thickness, in the center of the magnet.  Light pipes were used to
guide the infrared light from spectrometer to the sample, and the
transmitted light was detected with a bolometer operating at 1.2K
temperature. The light propagated parallel to the static magnetic
field.  Spectra were recorded at fixed fields. The frequency
resolution was selected at 0.2~cm$^{-1}$, much less than the width
of the resonance lines. The upper frequency cut-off was adjusted
to about 20~cm$^{-1}$ with a Fluorogold filter of appropriate
thickness. The lower cut-off was limited by the incident spectrum
and the spectrometer performance. The signal to noise ratio was
good down to 3~cm$^{-1}$, but there were strong minima in the
incident light intensity around 3.7 and 4.6~cm$^{-1}$. We
eliminated the experimental points at these minima without loosing
the data around them, obtaining reliable signal down to
3~cm$^{-1}$. After averaging over several spectra (typically
between two and four) and smoothing, the absorption spectrum  at a
given field was divided by a reference spectrum recorded at zero
magnetic field, following the same procedure as described  in Ref.
\cite{Mihaly LaMnO3}.

\section{Results}

First, we present the ESR results obtained with the classical ESR
technique. A typical spectrum (the derivative of the absorbtion
signal), recorded at 2.4~K at 75~GHz in the antiferromagnetic
phase, has three resonance features (Fig. \ref{Figure 1}). The
lines labeled (A), (B), and (C) are shifted from the paramagnetic
$g=2$ line differently and they exhibit different temperature
dependencies. Mode (A) has been already studied in \cite{NaNiO2
AF} and was related to the spin flop field at 1.8 T. Mode (B)
remains close to the paramagnetic line. Mode (C) consists of two
close resonances which broaden with temperature and are hardly
resolved above 6~K. Therefore only the first, larger intensity,
resonance is plotted for the temperature dependence of mode (C).
Below the phase transition temperature T$_{\rm N}$, this mode
behaves as the order parameter.

The paramagnetic line splitting at 200~K (empty symbols in Fig. 1)
is in agreement with the line splitting studied in detail at
several frequencies by Chappel \textit{et al.} \cite{NaNiO2 ferro
orbital}. There is also an agreement in the temperature dependence
of the splitting.  We will discuss the paramagnetic state in a
separate publication \cite{deBrion}.

The measurements were done on powder samples, and therefore the
large line broadening seen in Fig. \ref{Figure 1} is not
surprising. For any given crystallite in the sample, the line
position depends on the relative direction of external field and
the crystallographic axes. Relatively sharp features appear in the
spectrum because in the powder sample certain resonance
frequencies ({\textit e.g.} the highest and the lowest possible
values) acquire large statistical weight. This effect is well
known from the powder samples with significant $g$-factor
anisotropy.

It is also important to notice that the presence of three lines in
the spectrum at finite field does not necessarily mean that the
microscopic Hamiltonian of the system has three modes, for the same
reasons as above.  On the other hand, the number of modes seen at
zero external field (where all crystallites are equivalent) are
directly relevant to the Hamiltonian, as we will discuss later.

The low temperature ESR was further explored by far infrared
spectroscopic methods.  The advantage of detecting ESR by far-IR
spectroscopy is that one can readily map the power absorption over
the full range of magnetic fields and frequencies. The measurement
does not rely on sweeping the magnetic field; this becomes
especially important when one tries to discern features at zero
field, or compare ESR to inelastic neutron scattering done in zero
field.

\begin{figure}
  \includegraphics[width=7cm]{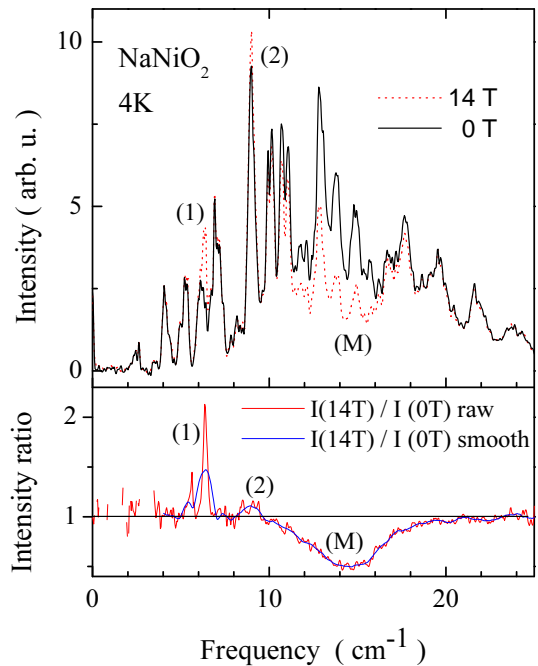}
  \caption{upper part: Typical FT-ESR spectra recorded at 4 K, 0~T and 14~T. Lower part: Intensity ratio.}
  \label{Figure 4}
\end{figure}

Typical absorbtion spectra at 0~T and 14~T recorded at 4~K are
given in Fig. \ref{Figure 4}. The sharp peaks in the spectra are
"instrumental": either due to interference features in the
synchrotron source, or generated by the multiple reflections in
the sample, or in the spectrometer itself. Nevertheless, the
absorption due to spins is clearly visible with features around
6~cm$^{-1}$ (labeled (1)), 9~cm$^{-1}$ (labeled (2)) and
11-18~cm$^{-1}$ (labeled (M)).

\begin{figure}
  \includegraphics[width=7cm]{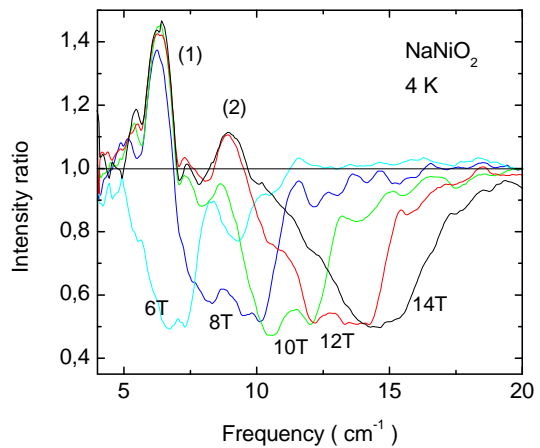}
  \caption{FT-ESR Relative transmission I$_{\rm r}(H)$ as a
function of the electromagnetic wave frequency at different
magnetic field values H.}
  \label{Figure 5}
\end{figure}

The key to these measurements is taking the ratio of two data sets
at different fields, eliminating the features in the raw spectrum
of the incident light, and leaving field dependent part of the
absorption. We have chosen always the same reference: the spectrum
at zero field, so that the relative transmission $I_{\rm r}$ is
related to the sample absorption $I_{\rm spin}(H)$:
\begin{equation}
I_r(H)=I_{\rm N}(H)/I_{\rm spin}(0).
\end{equation}
 The relative transmission for $H$=14~T is plotted in the bottom
part of fig. \ref{Figure 4}. The data below 3~cm$^{-1}$ are not
reliable. A smoothing procedure using a FT filter has been used to
improve the signal to noise ratio. It is responsible for the weak
oscillations with the 2~cm$^{-1}$ periodicity.
 The relative transmission data for several magnetic fields are plotted in
Fig. \ref{Figure 5} for the frequency range 4~cm$^{-1}$-20
cm$^{-1}$. An average over 3 to 4 data has been used. Below 10
cm$^{-1}$ two broad peaks ((1) and (2)) lie above 1, indicating
that the absorption $I_{\rm spin}(0)$ presents two resonance
modes. From all these relative transmission data, we can deduce
the frequency dependence of $I_{\rm spin}(0)$ and then $I_{\rm
spin}(H)=I_r(H)$$\times$$I_{\rm spin}(0)$ for each magnetic field
value.

\begin{figure}
  \includegraphics[width=8.5cm]{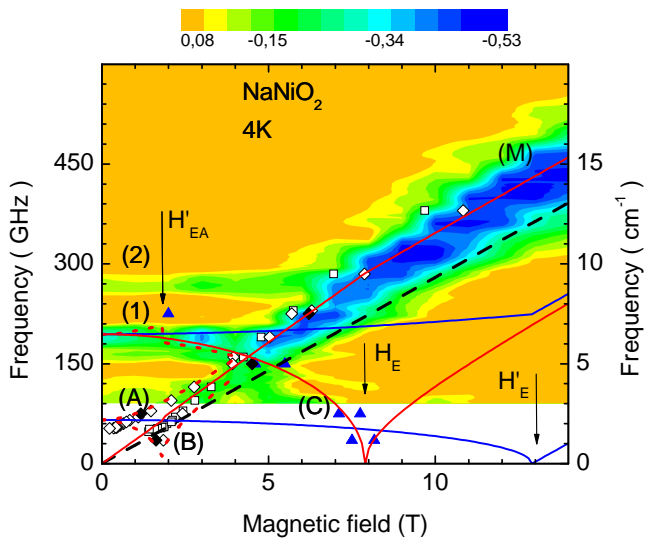}
  \caption{Magnetic modes at 4~K in a frequency versus field map
obtained from the resonance lines positions in conventional
ESR(closed symbols: this work; open symbols: from \cite{NaNiO2
AF}) and the transmission $I_{\rm spin}$ in FT-ESR (the color
scale on the top refers to the intensity of the signal, as shown
in Fig. \ref{Figure 5}). The dash black line corresponds to the
paramagnetic resonance at $g=2$. The blue and red continuous lines
represent the calculated frequencies $\omega^\parallel$ and
$\omega^\perp$ corresponding to fields parallel and perpendicular
to the hard axis, respectively. the red dotted lines,
$\omega^{int}$, are for fields along the intermediate axis (see
appendix B).}
     \label{Figure 3}
     \end{figure}

All these results can be mapped on a frequency / magnetic field
diagram (Fig. \ref{Figure 3}).  The symbols correspond to maxima
in the derivative of the ESR absorption measured by field sweep at
fixed frequencies, the colors are derived from the absorption
$I_{\rm spin}(H)$ measured at fixed fields. The main branch (M) is
a broad signal, with frequency approximately proportional to the
field, significantly above the $g=2$ line. Mode (A) and (B) are
related to the spin-flop transition observed at 1.8~T in the
magnetization curve \cite{NaNiO2 Bongers}. Mode (A) has been
discussed previously \cite{NaNiO2 AF} in terms of an easy axis
antiferromagnet model \cite{NaNiO2 AF}. The field dependence of
this mode is typical of crystallites in the powder where the
static field is perpendicular to the easy axis. When the magnetic
field is along the easy axis, a spin-flop transition occurs at
1.8~T. Crystallites oriented in this direction contribute to Mode
(B).

Modes (1) and (2) are seen mostly in the far-IR measurements, with
supporting evidence from a field sweep at 225 GHz.  These modes
have very little field dependence.  The corresponding zero field
gaps are 6.5 and 9.0~cm$^{-1}$, respectively (0.85 and 1.1~meV).
Mode (1) at 0.85~meV is close to the dispersionless magnon mode
observed at 0.7~meV in inelastic neutron scattering measurements
\cite{lewis2005}. Mode (1) crosses the main branch, and it appears
again at low frequencies, labelled as mode (C), also seen in Fig.
\ref{Figure 1}. Apparently, this mode softens to zero frequency
around 8~T.

\begin{figure}
  \includegraphics[width=8cm]{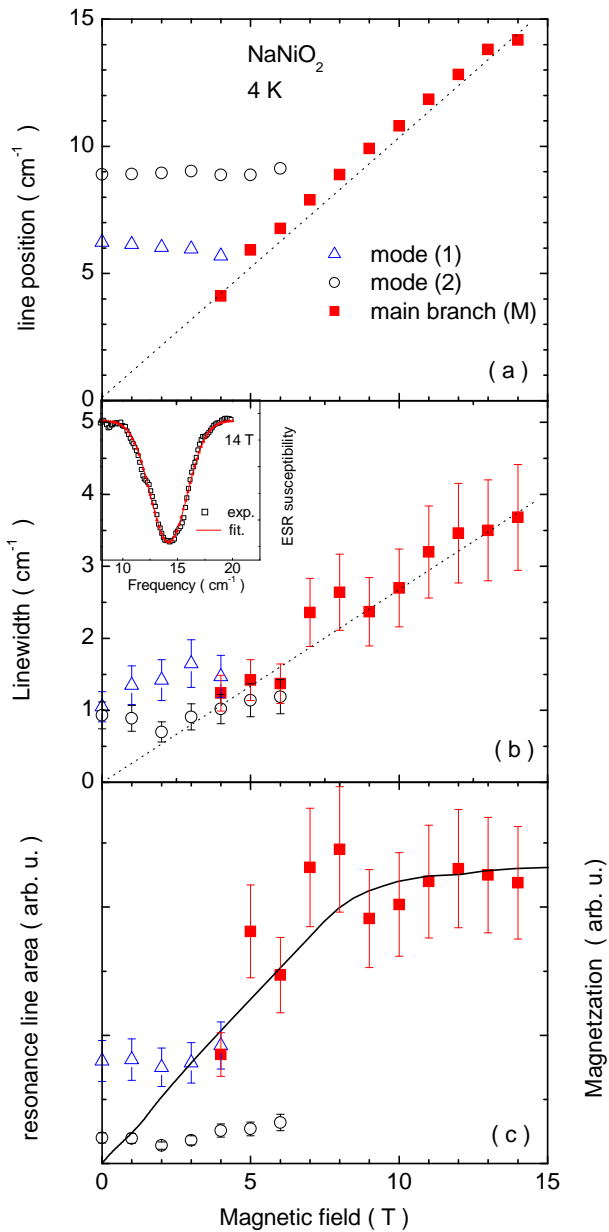}
  \caption{Gaussian fit of FT-ESR susceptibility at 4 K:
  line position (a), line width (b)and line total area (c)
    together with the sample magnetization (continuous line) as a function of applied magnetic
    field. The dotted line in (a) corresponds to the paramagnetic
    line at $g$=2.2 and in (b) to a $g$ anisotropy linewidth
    $\Delta$$g$=0.26. Insert: gaussian fit of the ESR susceptibility.}\label{gaussian}
\end{figure}

Further information can be deduced from a quantitative analysis of
the IR absorption data. We will assume that the absorbtion $I_{\rm
spin}(H)$ can be approximately related to the dissipative spin
susceptibility $\chi"$:
\begin{equation} \ln{I_{\rm spin}(H)}=\frac{-4\pi nd}{c}\chi"(H)
\end{equation}
where $c$ is the vacuum speed of light, $n$ the refraction index,
$d$ is thickness of the sample and $\omega$ the light frequency.
The absorption peaks $\chi"(H)$ are well fitted with a Gaussian
lineshape (see insert in Fig. \ref{gaussian}b). The corresponding
resonance frequency, linewidth and total area are plotted as a
function of the applied magnetic field (Fig.\ref{gaussian}a, b
,and c respectively). Note that the line positions are unchanged
whether one looks at $I_{\rm spin}(H)$ or $\chi"(H)$.  The line
position and line width confirm that the resonance associated with
the main branch (M) is nearly paramagnetic with $g \approx 2.2$.
Its linewidth is in accordance with the linewidth calculated from
the $g$ anisotropy at 200 K. However the anisotropic line shape
observed at 200~K is no longer present here, presumably due to
additional lifetime broadening. The branches (1) and (2) are
clearly identified at 6.5~cm$^{-1}$ and 9~cm$^{-1}$ with no field
dependence until they both merge into the main branch (M).

The area under the absorption peak is related  to the ESR
susceptibility through the Kramers-Kronig relation:
\begin{equation}
\chi(H,0)\propto \int \frac{\chi''(H,\omega)}{\omega}
d\omega\simeq\frac{1}{\omega_0}\int \chi''(H,\omega)d\omega
\end{equation}
 where $\omega_0$ is the peak
position. For the main branch (M), $\omega_0$ is proportional to
the static field H. Then the peak area $\int
\chi''(H,\omega)d\omega$ scales with the static magnetization.
This is indeed observed in Fig. \ref{gaussian}c. The saturation of
the total intensity above 9~T is in accordance with the saturation
of the static magnetization observed in this material
\cite{LiNaNiO}. For branches (1) and (2), $\omega_0$ is constant.
Then the peak area is proportional to the static susceptibility
$\chi(H,0)$. For both modes, it is constant with the field, as is
expected away from spin-flop field transitions. One can also see
that mode (2) has  a weaker susceptibility than mode (1). Note
that this susceptibility measured on a powder sample includes the
statistical weight associated with the crystallites in the powder
which are oriented correctly as regards the magnetic field
direction for each resonance mode.

\section{Discussion}
Several models have been proposed to describe the magnetic
properties in this material. The first one assumes a uniaxial easy
axis antiferromagnet, proposed  by Bongers \emph{et al.}
\cite{NaNiO2 Bongers} to account for the spin-flop transition at
1.8 T and then used in Ref. \cite{NaNiO2 AF} for the ESR magnon
branch at 53 GHz (branch (A) in Fig. \ref{Figure 3}). This model
includes a strong ferromagnetic coupling in the Ni planes, a
weaker antiferromagnetic coupling between the Ni planes and an
even weaker anisotropy which aligns the spins in the easy axis
direction. It was shown later that this model was incomplete in
terms of explaining the magnetic properties, and an additional
energy scale was required \cite{LiNaNiO}. This conclusion was
confirmed by recent neutron inelastic scattering measurements
\cite{lewis2005}. The magnon spectrum was interpreted in terms of
an easy plane AF model with the same ferromagnetic and
antiferromagnetic couplings as in the previous model.  The
anisotropy was assigned to the ferromagnetic exchange coupling,
and its magnitude was an order of magnitude larger than the weak
anisotropy used in the evaluation of the ESR data.

These models were merged by assuming an anisotropy tensor that has
an easy axis within an easy plane, with a much larger energy
needed for turning the spins out of the plane than within the
plane. This simple approach accounts for most of the experimental
observations, including a better description of the low field ESR
data of Chappel et al.\cite{NaNiO2 AF}. We start with the
following  anisotropic spin Hamiltonian:

\begin{eqnarray}
\label{Equa0} {\cal H}_0=
 \sum_{ab}{\bf S}_i \overleftrightarrow{J}_{ab} {\bf S}_j +
 \sum_{c}{\bf S}_i \overleftrightarrow{J}_c {\bf S}_j +
 g \mu_B{\bf S H}
\label{hamiltonian}
\end{eqnarray}
where ${\bf H}$ is the magnetic field, $\sum_{ab}$ is over nearest
neighbor spins within the layers, and $\sum_{c}$ is over nearest
neighbor spins along the $c$ directions. The exchange interactions
are represented by tensors: $\overleftrightarrow{J}_{ab}$
describes the ferromagnetic intraplane exchange, and
$\overleftrightarrow{J}_{c}$ is the interplane antiferromagnetic
exchange.  The intensity and the frequency of the spin resonance
lines for a crystal positioned in arbitrary direction relative to
the external field can be calculated from the Hamiltonian in two
steps.  First, the spin configuration is determined by minimizing
the free energy; second the ESR frequencies are related to the
small oscillations around the equilibrium configuration. There are
two resonance modes for any given direction when the magnetic
order is described by two sublattices. Occasionally (in zero field
or field applied along high-symmetry directions) the two modes may
be degenerate.

The powder spectrum is obtained by the average of the absorption
of individual grains over all directions.  In practice, the
calculation of frequency and intensity for arbitrary field
orientation is rather hard. Instead, the resonance frequency is
calculated for the static field pointing in three principal
directions. The six "principal frequencies" (three frequencies for
each of the two modes) play a special role in the powder average,
since some of them correspond to extremal values and the
statistical weight factor for these frequencies will be high.

This process  is similar to the determination of the
$\overleftrightarrow g$ tensor from the paramagnetic resonance of
a powder sample with a $g$-factor anisotropy. Naturally, the
orientation of the principal axes of the tensor relative to the
crystallographic directions cannot be determined from measurements
on a powder sample. Nevertheless, the $\overleftrightarrow
g$-tensor components are routinely determined from powder
measurements this way.  It is often sufficient to look at the
highest and lowest frequencies in the measured broad ESR line, and
one can identify the highest and lowest tensor components.

There is an extensive literature on the anisotropy effects in
antiferromagnets, starting with the early work of Keffer and
Kittel \cite{keffer} on the easy axis problem.  The three axis
case (easy, intermediate, hard) was treated by Nagamiya
\cite{nagamiya}. Here we base our analysis on this work and on a
review by Turov \cite{turov}. The magnon spectrum is treated in
the linear approximation, equivalent to a quasi-classical
treatment and we focus on the ground state properties ($T$=0~K).
The details of the calculation are given in Appendix A and B.
First, the Hamiltonian is transformed into:

\begin{equation}
\label{turov1}
 {\cal H}_{MF}/N= {A \over 2} {\bf m} ^2 + {a\over 2} m_z^2 + {b \over 2} l_z^2
 +{c\over 2} m_x^2 + {d \over 2} l_x^2 -{\bf m} {\bf h},
\end{equation}
where ${\bf m}$ and ${\bf l}$ are the total magnetization and the
antiferromagnetic order parameter, respectively. Parameter $A$
describes an isotropic antiferromagnetic coupling,  $a$ and $c$
describe the anisotropy of the total ferromagnetic moment. In the
ideal antiferromagnetic state ${\bf m}=0$, therefore these
parameters have no influence on the free energy of the system in
zero external magnetic field.  On the other hand, parameters $b$
and $d$ act on the order parameter $\bf l$ directly. Kittel
treated the uniaxial model ($c=d=0$) at low fields; easy plane and
easy axis corresponds to $b > 0$ and $b < 0$, respectively.
Nagamiya's biaxial model is obtained when $a=c=0$. Turov discusses
the uniaxial case for fields up to the saturation field
\cite{turov}.

We set $b > 0$ so that the zero field equilibrium magnetizations
are in the plane perpendicular to $z$; this will be the "easy
plane". The anisotropy within the plane, represented here by
coefficients $c$ and $d$, is much smaller: $c,d<<b$. (Notice that
these choices do not follow from the symmetry arguments related to
the lattice distortion, but they were forced by the experimental
observations). Since $c$ and $d$ have similar effects on the spin
resonance frequencies, and we can not determine them separately,
we will assume that $d=0$. This leaves us with four parameters to
determine:  $A$, $a$, $b$ and $c$.  For $c<0$, $x$ will be the
easy direction, $y$ will be the intermediate direction, and $z$
remains the hard direction.

To facilitate the discussion, it is convenient to introduce the
"effective fields" $H_E=A/M_0$, $H_E'=(A+a)/M_0$, and
$H_{EA}'=\sqrt{Ac}/M_0$  where $M_0$ is the saturation
magnetization (more effective fields follow in Appendix B). For
applied field perpendicular to the hard axis the spin system
saturates at $H_E$, when the external field overcomes the exchange
field.  For applied field parallel to the hard axis the saturation
occurs at $H_E'$ \cite{turov}. There is a spin-flop transition for
field along the easy axis at $H=H_{EA}'$ \cite{nagamiya}.

The published susceptibility data, Fig. 4 of Ref.
\onlinecite{LiNaNiO} can be used to find all but one of the
parameters of the model.  The first peak in the $dM/dH$ curve at
1.8~T has been already identified as the spin-flop field $H_{EA}'$
\cite{NaNiO2 AF}. At around 8~T the $dM/dH$ curve has a shoulder. We
identify this as the onset of the saturation, and the corresponding
field is $H_E$. Finally, the saturation is complete around 13T,
corresponding $H_E'$.  From these values we get $A=8~T$, $a=5~T$ and
$c=-0.42~T$.

The parameter $b$ determines the ESR frequency $\omega_2$ at zero
field. We selected $b=6$~T  to match the zero field gap at
6.5~cm$^{-1}$ (Mode (1) in Figures \ref{Figure 3} and
\ref{gaussian}. This is quite close to the dispersionless magnon
mode seen in neutron scattering (0.7~meV = 5.6~cm$^{-1}$)
\cite{lewis2005}.

The principal frequencies are listed in Appendix B, and the results
are shown in Fig.  \ref{Figure 3}.  Considering that many of the
parameters of the model were taken from other measurements, the
agreement with the experiment is excellent. The calculated zero
field resonance $\omega_1^{hard}$ is very close to the measured one
at 53~GHz \cite{NaNiO2 AF}. In fact, the low field behavior of our
model (where the slope $d\omega / d H$ is zero for $H \rightarrow
0$) fits the lower branch in Fig. 8 of Ref. \onlinecite{NaNiO2 AF}
better than the uniaxial easy axis model used in that work (where
the $d\omega / d H=-\gamma$). Mode (1) is identified with
$\omega_2$. As the field is increased, this mode is approximately
independent of the field for any field direction.  Therefore the
powder signal remains narrow, and it is clearly visible in the
fixed-field far-IR spectroscopy scans.

Field-independent modes, like the $\omega_2^{hard}$ and
$\omega_2^{int}$ modes here, are very hard to detect in the
field-sweep scan commonly used with the fixed-frequency methods.
However, the spin-flop transition that occurs in the part of the
sample where the field is close to the easy direction causes a
jump in the frequency of $\omega_2^{easy}$. This jump creates a
measurable signal in the field-sweep scan \cite{nagamiya}. In our
case the 225~GHz fixed frequency ESR measurement was just in the
right frequency range to catch this feature.

Above 5~T the powder average signal originating from $\omega_2$
broadens and merges with other absorption. It can not be seen in the
far-IR measurement, and only the (more sensitive) fixed frequency
study picks it up as Mode (C). The upper extremum of all possible
frequencies, $\omega_1^\perp$, yields a strong signal. Although the
$g$-factor was not adjusted ($g=2$), the coefficient $\sqrt
{1+H_a/H_E}$ seen in Eq. \ref{Equa3} brings $\omega_1^\perp$ right
into the middle of the experimental values of the main branch (M).

With the set of parameters $A$=8~T, $a$=5~T, $b$=6~T and
$c$=~-0.42~T, we can deduce the coupling energies in the
Hamiltonian described by Eq. \ref{Equa0}. For the
antiferromagnetic coupling we get $J_{AF}^z=A/3$=3.58~K. The
anisotropy of the ferromagnetic interaction is
$J_F^z-J_F^y=(a+b)/3$=4.93~K, comparable to the antiferromagnetic
coupling. The anisotropy of the antiferromagnetic coupling is weak
$J_{AF}^z-J_{AF}^y=(a-b)/3$=0.45~K. The ferromagnetic coupling,
deduced from the Curie-Weiss temperature, $J_F^z$=-15~K, is the
dominant interaction. This hierarchy of interactions was also
deduced from inelastic neutron scattering by Lewis \textit{et
al.}\cite{lewis2005} using a similar model.

The weak anisotropy within the easy plane can be assigned to the
ferromagnetic or antiferromagnetic couplings or a combination of
the two.  The parameter $c$ represents the sum of the two
anisotropies:
$2c/3=(J_{F}^x-J_{F}^z)-(J_{F}^y-J_{F}^z)+(J_{AF}^x-J_{AF}^z)-(J_{AF}^y-J_{AF}^z)=-0.37$~K.
However, the measurements would be compatible with other
combinations of these parameters (as long as the sum is fixed).
For example all anisotropy may be in the ferromagnetic coupling
$(J_{F}^x-J_{F}^z)-(J_{F}^y-J_{F}^z)=-0.37~$K and no anisotropy in
the antiferromagnetic coupling,
$(J_{AF}^x-J_{AF}^z)-(J_{AF}^y-J_{AF}^z)=0$.

We tested the model with various other choice of parameters,
partially disregarding the three characteristic fields obtained
from the susceptibility measurements. Decreasing $a$ to zero fails
to describe the flatness of Mode (1). Negative values of $a$ do
not give good fits.  For $a=b$ the agreement with the data is
still reasonable; in this special case the uniaxial anisotropy of
the antiferromagnetic interplane coupling, $J_{AF}^z-J_{AF}^y$, is
exactly zero.

In spite of the general agreement between the model and the
experiment, there are two features that remain unexplained: the
splitting of mode (C) at low temperature (Figure \ref{Figure 1},
insert) and the existence of mode (2).

The splitting if mode (C) is consistent with our model. This mode is
detected when the field is perpendicular to the hard axis. Note that
the calculations for $\omega_2^\perp$ were done with $c=0$, but in
the presence of a finite $c$ the saturation field may be slightly
different for static fields along the easy and intermediate axes.
The magnitude of parameter $|c|=0.42~$T is in the right range to
explain the observed splitting.

The existence of mode (2) is entirely beyond the simple model
presented here.  The two-sublattice Hamiltonian yields two modes at
zero field, and we have those two modes identified as mode (A) at
1.75 cm$^{-1}$ (53 GHz) and mode (1) at 6.5 cm$^{-1}$.  No room is
left for mode (2) in this model.

\section{Conclusion}

We studied the field dependence of the ESR in NaNiO$_2$ and we
interpreted the results in terms of a model involving anisotropic
exchange interactions. The main interaction is ferromagnetic in
the triangular NiO plane. It is an order of magnitude larger than
the interplane antiferromagnetic coupling. The anisotropy in the
ferromagnetic coupling is characterized by an uniaxial "easy
plane" parameter of $a+b=11$~T, and a much weaker easy axis
parameter of $c=-0.42$~T.

One of the main results of our work is that the easy direction is
only slightly better than the other directions perpendicular to
the hard axis.  This is in general agreement with the inelastic
neutron scattering result, where the magnon spectrum was modeled
with an easy plane spin Hamiltonian \cite{lewis2005}.  For the
easy plane system  one of the magnon modes has zero energy (in
zero field and at zero wavenumber).  The anisotropy within the
easy plane pushes this magnon mode to a finite energy
(1.8~cm$^{-1}$ = 0.2~meV). This second magnon gap is small, and
therefore it is not visible in the neutron scattering, but it is
clearly seen in the ESR studies.

\begin{figure}
\includegraphics[width=7cm]{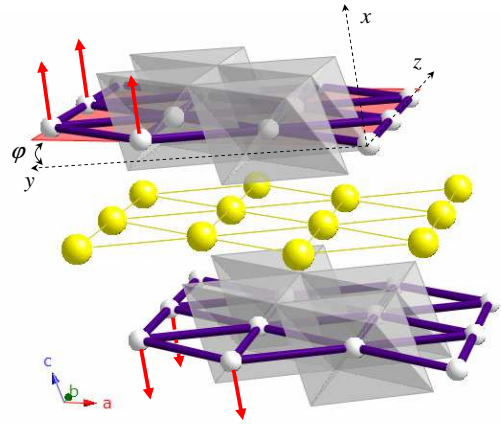}
\caption{The orientation of the hard axis $z$ and the easy axis $x$
relative to the triangular Ni layers.  The $ab$ plane is highlighted
in the top layer.  The $z$ axis is parallel to the $b$ direction in
the crystal; the angle between the $ab$ plane and the $y$ axis is
$\varphi \approx 10^{\circ}$.  The spin directions are indicated by
red arrows.}
 \label{Figure 9}
\end{figure}

The principal axes of the anisotropy tensor were constrained by
symmetry considerations (see appendix A), but cannot be determined
uniquely. In order to reconcile our results with the magnetization
measurements on a twinned crystal \cite{NaNiO2 Bongers} and with the
neutron diffraction results \cite{NaNiO2 neutron}, where the spin
orientation was measured, the easy direction (the $x$ axis of the
reference frame) was selected so that it makes a $~100^{\circ}$
angle to the $ab$ plane (see Fig. \ref{Figure 9}). The hard axis was
selected within the triangular nickel plane, along the $b$ direction
of the crystal (Fig. \ref{Figure 9}).  Alternatively, it may point
in the $y$ direction of the reference frame.  The hard axis is
probably related to spin-orbit effects on the ferromagnetic
exchange. Indeed, the ratio $|(J_F^z-J_F^y/J_F^y|=0.33$ is
comparable to the spin-orbit effect on the $g$ tensor:
($g^\perp-g^\parallel)/g^\parallel=0.10$. The choice of the easy
axis involves an energy scale one order of magnitude smaller.

The existence of a a third zero-field mode (mode 2) may point
beyond the model used here: any two-sublattice model, that assumes
that the magnitude of the sublattice magnetisations are fixed,
yields at most two resonance modes in zero field. One can then
think of quantum effects. They may cause some modes to shift from
zero to finite frequencies \cite{Chubukov}, but do not change the
number of modes. The comparison with other triangular magnets
having antiferromagnetic coupling {\it within} the layers is
intriguing: (including CsNiBr$_3$, RbNiBr$_3$, CsMnI$_3$,
CsMnBr$_3$ CsCuCl$_3$, RbCuCl$_3$ and RbFe(MoO$_4$)$_2$). There,
typically three or six ESR modes \cite{tanaka, Chubukov, Smirnov}
are observed. This is related to the umbrella-like spin order
with, for a given triangle, each spin pointing at 120 $^{\circ}$
from each other. Three sublattices are definitely present. This is
not supposed to be the case in NaNiO$_2$ where there is strong
support that the  coupling within the triangular layers is
ferromagnetic (the Curie-Weiss temperature is positive and the
neutron diffraction patterns correspond to a A type
antiferromagnet with ferromagnetically aligned spins in the layers
\cite{NaNiO2 neutron}). However the sign of the interactions in
the triangular planes has been the subject of intense debate in
the isomorphic compound LiNiO$_2$ \cite{LiNiO2 exp}. Even for
NaNiO$_2$ a recent calculation by Vernay \textit{et al}
\cite{LiNiO2 theo} proposes that one out of three interactions
bond in the triangles are antiferromagnetic leading to
antiferromagnetic chains, although this magnetic structure has not
been observed experimentally. More over, in the intermediate
compound Li$_{0.3}$Na$_{0.7}$NiO$_2$ \cite{LiNaNiO}, which, on the
orbital sector, behaves as LiNiO$_2$ and, on the spin sector,
presents a long range antiferromagnetic order as in NaNiO$_2$, we
also observe the occurrence of three modes at zero field (around
0, 5.2 and 6.5~$cm^{-1}$); there the magnetic structure has not
been solved yet. It is tempting to assign these similar behaviors
to the presence of additional antiferromagnetic interactions.
Further studies, preferably on single crystals, may solve this
remaining puzzle in the NaNiO$_2$/LiNiO$_2$ layered triangular
compounds.

\section{Acknowledgement}
We are indebted to G.L. Carr for valuable discussions and for
developing the IR facilities at the NSLS, F. Mila, F. Vernay, C.
Lacroix and M.D.  N\'{u}\~{n}ez-Regueiro for their theoretical
inputs. S. de Brion acknowledges the Universit\'e Joseph Fourier
for a CRTC program. The Budapest authors acknowledge the Hungarian
State Grants (OTKA) TS049881, T60984 and F61733, and the
MERG-CT-2005-022103 EU Project. F.S. acknowledges the Magyary
program for support, L.M. acknowledges the Szent-Gy\"orgyi
fellowship. The Grenoble High Magnetic Field Laboratory is
"laboratoire Conventionn\'{e}e \`{a} L'Universit\'{e} Joseph
Fourier". Use of the National Synchrotron Light Source, Brookhaven
National Laboratory, was supported by the U.S. Department of
Energy, Office of Science, Office of Basic Energy Sciences, under
contract No. DE-AC02-98CH10886.

\section{Appendix A: derivation of the mean field Hamiltonian and symmetries}
In the molecular field approximation the spins are assigned to two
sublattices: Spins in the even layers are labeled by $\mu$ and spins
in the odd layers are labeled by $\nu$.  ${\bf S}_i$ and ${\bf S}_j$
in the Hamiltonian are replaced with their average values, ${\bf
S}_1=2/N \sum {\bf S}_\mu$ and ${\bf S}_2=2/N \sum {\bf S}_\nu$,
where N is the total number of spins.

First, let us look at the term $\sum_{ab}{\bf S}_i
\overleftrightarrow{J}_{ab} {\bf S}_j $ in the Hamiltonian Eq.
\ref{hamiltonian}.  All of the spins in this sum belong to the
same sublattice, so the terms to add up will have the form of
${\bf S}_1 \overleftrightarrow{J}_{ab} {\bf S}_1 $ and ${\bf S}_2
\overleftrightarrow{J}_{ab} {\bf S}_2 $.  For any given spin
within a ferromagnetic layer, there are six neighbors,
approximately positioned on the corners of a hexagon. Even though
the exchange $\overleftrightarrow{J}_{ab}$ with any one of this
neighbors may be anisotropic, in the sum of the six terms most of
the anisotropy within the plane will be cancelled.  In fact, as
long as the exchange couplings can be represented by tensors, in
an undistorted lattice the cancellation will be exact.  The
corresponding part of the Hamiltonian would be: $\sum_{ab}{\bf
S}_i \overleftrightarrow{J}_{ab} {\bf S}_j = (N/2) (1/2) 6({\bf
S}_1 \overleftrightarrow{J_F} {\bf S}_1+{\bf
S}_2\overleftrightarrow{J_F}{\bf S}_2)$, where the effective
coupling $\overleftrightarrow {J_F}$ is isotropic within the
plane, the extra factor of 1/2 compensates for the double counting
of each bond and the factor 6 reminds us that this exchange is an
average over six bonds.  Similar arguments can be used for the
antiferromagnetic interlayer coupling to derive
$\overleftrightarrow{J}_{AF}$ from a sum over
$\overleftrightarrow{J}_{c}$'s.   The proper reference frame for
the new coupling tensors has the $x$ direction perpendicular to
the layers. The (equivalent) $y$ and $z$ directions are within the
layers.

The anisotropy of the effective couplings
$\overleftrightarrow{J}_{F}$ and $\overleftrightarrow{J}_{AF}$ is
related to the distortion of the lattice.   While above 480~K all
nearest neighbor Ni-Ni bond lengths are equal \cite{NaNiO2 ferro
orbital}, below the transition one side of the triangular lattice
(along the $b$ direction) is compressed, and two other sides are
elongated. The $a-c$ plane remains a mirror plane, but the direction
perpendicular to the layers is not a high symmetry direction
anymore. (See Fig. \ref{Figure 9}.) Accordingly, the $x$ direction
will tilt relative to the direction perpendicular to the layers, as
it acquires a non-zero component in the crystallographic $a$
direction. At the same time the anisotropy axes within the plane
become well defined. We will take the $z$ axis parallel to the $b$
crystallographic direction. The $y$ axis will point slightly out of
the plane, being perpendicular $x$ and $z$. Notice that, except for
permutations of the  $x,y,z$ labels, this is the only selection
satisfying the broken crystal symmetries. The magnitude of the tilt
of the $x$ axis, and the amount of anisotropy, remain a free
parameters. Similar arguments apply to the nearest neighbor
antiferromagnetic coupling between the layers.

Spin resonance on a powder sample can be used to evaluate the
anisotropy parameters of the resonance line.  As we will see
later, our system is described by a model with a dominant "easy
plane" anisotropy, and a much weaker anisotropy within the plane.
However, the orientation of the anisotropy axes relative to the
crystal cannot be can not be determined from a powder measurement.
Neutron diffraction \cite{NaNiO2 neutron} and single crystal
magnetization data \cite{NaNiO2 Bongers} indicate that the easy
axis makes an angle of approximately $100^{\circ}$ to the NiO
planes.  We select this to coincide with the easy direction (the
$x$ axis of the reference frame). The hard axis may be either
along $y$ or $z$; both choices agree with existing symmetries. We
tentatively pick the $z$ direction for the hard axis.

The mean field Hamiltonian becomes:
\begin{widetext}
\begin{eqnarray}
\label{meanfield}
 {\cal H}_{MF}=
 {3N \over 2}\left[
 J_F^x (S_{1x}^2+S_{2x}^2)
 +J_F^y (S_{1y}^2+S_{2y}^2)
 +J_F^z (S_{1z}^2+S_{2z}^2)\right]+
 \nonumber\\
 +3N \left(
 J_{AF}^x S_{1x} S_{2x}
 +J_{AF}^y S_{1y} S_{2y}^2
 +J_{AF}^z S_{1z} S_{2z}^2
 \right)+
 {N \over 2}g \mu_B ({\bf S}_{1}+{\bf S}_{2}){\bf H}
\end{eqnarray}
The dimensionless sublattice magnetizations are defined as ${\bf
m}_1= {1 \over 2} {\bf S}_1/ S$ and ${\bf m}_2= {1 \over 2}{\bf
S}_2 / S$. When thermal and quantum fluctuations are neglected,
the length of these vectors is fixed $|{\bf m}_1|= |{\bf
m}_2|=1/2$. Accordingly, we can express one of the sublattice
magnetization components with the other two:
$m_{1z}^2=1/4-m_{1x}^2-m_{1y}^2$ and
$m_{2z}^2=1/4-m_{2x}^2-m_{2y}^2$. The Hamiltonian is:

\begin{eqnarray}
\label{meanfield1}
 {\cal H}_{MF}= 12NS^2 \left[
 (J_F^x - J_F^z)(m_{1x}^2+m_{2x}^2)
 +(J_F^y- J_F^z) (m_{1y}^2+m_{2y}^2)\right]+
 \nonumber\\
 +6NS^2 \left(
 J_{AF}^x m_{1x} m_{2x}
 +J_{AF}^y m_{1y} m_{2y}
 +J_{AF}^z m_{1z} m_{2z}
 \right)+
 {N S g \mu_B ({\bf m}_1}+{\bf m}_2){\bf H}
\end{eqnarray}
and we dropped a constant term proportional to $J_F^z$.  In the
absence of fluctuations the magnitude of the ferromagnetic coupling
becomes irrelevant, only the anisotropy matters. Interestingly, the
anisotropy term obtained here for the ferromagnetic coupling cannot
be distinguished (within the molecular field approximation) from a
microscopic "single ion" spin anisotropy. However, the physical
origins are different: For spin 1/2 the microscopic single ion
anisotropy can be shown to be exactly zero, whereas the same
arguments do not apply here, where the anisotropy is due to the
ferromagnetic exchange coupling.

The easy plane anisotropy around the $z$ axis corresponds to $(J_F^x
- J_F^z)\approx(J_F^y - J_F^z)$ (and similarly for the
antiferromagnetic coupling). We can re-arrange the terms to better
reflect this fact. For $m_1$ we get
\begin{eqnarray}
(J_F^x - J_F^z)m_{1x}^2+(J_F^y - J_F^z)m_{1y}^2=
 \nonumber\\
 =1/2[(J_F^x - J_F^z)+(J_F^y - J_F^z)](m_{1x}^2+m_{1y}^2) +
 1/2[(J_F^x - J_F^z)-(J_F^y - J_F^z)](m_{1x}^2-m_{1y}^2)=
 \nonumber\\
 =1/2[(J_F^x - J_F^z)+(J_F^y - J_F^z)](1-m_{1z}^2) +
 1/2[(J_F^x - J_F^z)-(J_F^y - J_F^z)](2m_{1x}^2-1+m_{1z}^2)=
 \nonumber\\
 =(J_F^z - J_F^y)m_{1z}^2 + [(J_F^x - J_F^z)-(J_F^y - J_F^z)]m_{1x}^2
 +const.
\end{eqnarray}
This way the coefficient of the $m_{1z}^2$ term represents the
uniaxial anisotropy; the (smaller) coefficient of the $m_{1x}^2$
term describes the remaining anisotropy around the $z$ axis. Similar
rearrangement can be made for $m_2$ and for all of the
antiferromagnetic terms.

The total magnetization and the antiferromagnetic order parameter
vectors are introduced as $\bf m= (\bf m_1 + \bf m_2)$ and $\bf l=
(\bf m_1 - \bf m_2)$.   Notice that $|{\bf m}|=1$ corresponds to the
saturation magnetization, and $|{\bf l}|=1$ describes perfect
antiferromagnetic order.  In terms of these new parameters the
Hamiltonian can be re-written as:

\begin{eqnarray}
\label{meanfield2}
 {\cal H}_{MF}= 3NS^2\left[
 J_{AF}^z({\bf m}^2-{\bf l}^2)
 +(J_{AF}^z - J_{AF}^y)(m_z^2-l_z^2)+
 (J_{AF}^x - J_{AF}^z)-(J_{AF}^y - J_{AF}^z)(m_x^2-l_x^2)
 \right.
 \nonumber\\ \left.
 +(J_F^z- J_F^y) (m_z^2+l_z^2)+
 +[(J_F^x - J_F^z)-(J_F^y - J_F^z)(m_x^2+l_x^2)\right]+
 {N}g \mu_B {\bf m}{\bf H}
\end{eqnarray}
Here we re-grouped the terms in the antiferromagnetic coupling,
emphasizing the anisotropy, without introducing new constrains.
Neglecting fluctuations leads to ${\bf m} {\bf l}=0$ and ${\bf
m}^2+{\bf l}^2=1$. Eliminating the ${\bf l}^ 2$ term finally yields:
\begin{equation}
\label{Equa1}
 {\cal H}_{MF}/N= {A \over 2} {\bf m} ^2 + {a\over 2} m_z^2 + {b \over 2} l_z^2
 +{c\over 2} m_x^2 + {d \over 2} l_x^2 -{\bf m} {\bf h}
\end{equation}
where $A=12S^2J_{AF}^z$,   ${ \bf h} = g \mu_B S {\bf H}$, and the
anisotropy constants are expressed as
\begin{eqnarray}
 a=6S^2(J_{F}^z-J_{F}^y+J_{AF}^z-J_{AF}^y){\rm , ~~~~~~~~}
 b=6S^2(J_{F}^z-J_{F}^y-J_{AF}^z+J_{AF}^y),
\nonumber \\
 c=6S^2[(J_{F}^x-J_{F}^z)-(J_{F}^y-J_{F}^z)+(J_{AF}^x-J_{AF}^z)-(J_{AF}^y-J_{AF}^z)]
\nonumber \\
d=6S^2[(J_{F}^x-J_{F}^z)-(J_{F}^y-J_{F}^z)-(J_{AF}^x-J_{AF}^z)+(J_{AF}^y-J_{AF}^z)],
\end{eqnarray}
Notice that Eq. \ref{Equa1} is applicable in the most general case,
but the coupling constants reflect an expected hierarchy.  A simple
isotropic antiferromagnet is obtained for $a=b=c=d=0$. For uniaxial
anisotropy $c=d=0$; the coefficients $a$ and $b$ describe
anisotropies in the total magnetization and antiferromagnetic order,
respectively.  In our case coefficients $c$ and $d$ will be small.
\end{widetext}

\section{Appendix B: spin resonance frequencies.}

For any given field direction there are two modes, denoted by
$\omega_1$ and  $\omega_2$.  At high external fields one can
neglect the small anisotropy within the easy plane, and take
$c=d=0$.  The frequencies were determined by Turov \cite{turov}.
There are two principal frequencies for each modes, corresponding
to external field applied parallel and perpendicular to the hard
axis. The results are expressed in terms of effective fields:
\begin{eqnarray}
\label{Equa2} && H_E=A/M_0
\nonumber\\
&&H_E'=(A+a)/M_0
\nonumber\\
&&H_E''=(A-b)/M_0
\nonumber\\
&&H_a=a/M_0
\nonumber\\
&&H_b=b/M_0
\nonumber\\
&&H_{EA}=\sqrt{H_E H_b}
\nonumber\\
&&H_{\parallel}= \sqrt {H_E'' H_b}
\nonumber\\
&&H_{\perp}=H_{\parallel}{H_E' \over H_E''}
\nonumber\\
\end{eqnarray}
where $M_0$ is the saturation magnetization.

For field perpendicular to the hard axis one gets:

\begin{eqnarray} \label{Equa3}
&&\omega_1^\perp = \gamma H \sqrt {1+H_a/H_E} \textrm{~~~~~if~~~~~}
H<H_E
\nonumber\\
&&\omega_1^\perp = \gamma \sqrt {H(H+H_a)}  \textrm{~~~~~if~~~~~}
H>H_E
\nonumber\\
&&\textrm {and ~~~~~ }
\nonumber\\
&&\omega_2^\perp = \gamma H_{EA} \sqrt {1- H^2/H_E^2}
\textrm{~~~~~if~~~~~} H<H_E
\nonumber\\
&&\omega_2^\perp = \gamma \sqrt {(H-H_E'')(H-H_E)}
\textrm{~~~~~if~~~~~} H>H_E
\end{eqnarray}
where $\gamma= g \mu_B /\hbar $ in terms of the Bohr magneton and
the $g$-factor.  $H_E$ is the saturation field; for $H>H_E$ all
spins line up with the external field.

For the field direction parallel to the hard axis the saturation
field is $H_E'>H_E$, since both the exchange coupling and the
anisotropy work against the external field. The frequencies are:

\begin{eqnarray}
&&\omega_1^\parallel = 0 \textrm{~~~~~if~~~~~} H<H_E'
\nonumber\\
&&\omega_1^\parallel = \gamma (H-H_E') \textrm{~~~~~if~~~~~} H>H_E'
\nonumber\\
&&\textrm {and ~~~~~ }\nonumber\\
&&\omega_2^\parallel = \gamma H_{EA} \sqrt {1+H^2/H_{\perp}^2}
\textrm{~~~~~if~~~~~} H<H_E'
\nonumber\\
&&\omega_2^\parallel = \gamma (H-H_a) \textrm{~~~~~if~~~~~} H>H_E'
 \label{Equa4}
\end{eqnarray}

The modes are shown in the high field part of Fig. \ref{Figure 4}.
The $\omega_2^\perp$ mode reaches zero at the lower saturation field
$H_E$.  Below the saturation field the $\omega_1^\perp$ mode looks
like a "free" spin resonance, except for the apparent $g$-factor is
increased by a factor of $\sqrt {1+H_a/H_E}$.  Well above the
saturation field the line is shifted to higher frequencies by the
amount of $\gamma H_a/2$, but its slope is still $\gamma$. For
fields parallel to the hard axis the $\omega_1^\parallel$ mode is
zero at the upper saturation field. With the particular choice of
parameters indicated in the Figure. The $\omega_2^\parallel$ mode is
nearly independent of the field up to  $H_E'$.

At low fields the anisotropy within the easy plane becomes
important.  A finite value of the parameter $c$ (or $d$) has two
important consequences. First, in zero field, the $\omega_1$ mode
(the "Goldstone mode") will shift to finite frequency. Second, the
field dependence of the $\omega_1^\perp$ mode will be different for
external fields applied in different directions within the easy
plane. Instead of the two principal directions relative to the hard
axis $z$ (parallel and perpendicular), we will have to deal with
three axes. We will identify these directions as the hard axis
(parallel to $z$), the intermediate axis and the easy axis (formerly
the two equivalent directions perpendicular to $z$).

For an approximate treatment of the $c \ne 0$ case we will turn to
an early paper by Nagamiya \cite{nagamiya}, assuming  $H \ll H_E$.

In finite fields along the hard axis the $\omega_1$ mode is
approximately independent of the field (this mode corresponds to the
rotation of the spins in the easy plane, and it is excited with an
oscillating magnetic field parallel to the static field). For fields
in the intermediate direction the frequency of the mode increases;
for the field applied in the easy direction there will be a
spin-flop transition at $H=H_{EA}'=\sqrt{Ac}/M_0$.

\begin{eqnarray}
\label{Equa5} &&\omega_1^{hard} = \gamma' H_{EA}' =
\omega_1^{\parallel }
\nonumber\\
&&\omega_1^{int} = \gamma' \sqrt{H^2+H_{EA}'^2}
\nonumber\\
&&\omega_1^{easy} = \gamma' \sqrt { {\alpha \over 2} + H^2 -\sqrt
{\beta^2/4+2\alpha H^2}}
\textrm{~~~if~~~}H<H_{EA}'\nonumber\\
&&\omega_1^{easy} = \gamma' \sqrt { H^2 -H_{EA}'^2}
\textrm{~~~~~if~~~~~} H>H_{EA}'
\end{eqnarray}

Here we used $\alpha = H_{EA}^2+H_{EA}'^2$ $\beta =
H_{EA}^2-H_{EA}'^2$. Nagamiya's calculation does not account for the
increase of the apparent $g$-factor seen in Turov's low field
result. Therefore we are using a modified $\gamma'= \gamma \sqrt
{1+H_a/H_E}$ here.

The anisotropy within the easy plane has no influence on the
$\omega_2^{//}$ mode (for fields parallel to the hard axis).  For
fields applied in the intermediate direction, the $\omega_2$ mode is
approximately independent of the field (except for a slight decrease
in frequency, a precursor of the eventual saturation at $H_E$). When
the field is in the easy direction, the spin-flop transition results
in a jump of the ESR frequency.

\begin{eqnarray}
\label{Equa6} &&\omega_2^{hard} = \gamma H_{EA}
\nonumber\\
&&\omega_2^{int} = \gamma H_{EA}
\nonumber\\
&&\omega_2^{easy} = \gamma \sqrt {{\alpha \over 2} + H^2 +\sqrt
{\beta^2/4+2\alpha H^2}}
\textrm{~~~if~~~}H<H_{EA}'\nonumber\\
&&\omega_2^{easy} = \gamma \sqrt { H_{EA}^2 -H_{EA}'^2}
\textrm{~~~~~if~~~~~} H>H_{EA}'
\end{eqnarray}

To ensure a seamless match to the high field calculation, in Fig.
\ref{Figure 3} we multiplied these frequencies with a saturation
term, $\sqrt{1-H^2/H_E^2}$.  For the hard axis direction we used the
original $\omega_2^{\parallel}$ values.  Neither this procedure, nor
the modified $\gamma'$ introduced earlier, can completely substitute
for the full solution of the problem for arbitrary fields and
parameters, but they are qualitatively correct.

\end{document}